\documentclass[12pt]{article}

\usepackage{epsfig}
\usepackage{rotating}

\begin{document}

\title{\Large \bf Two-Pion Production in Nucleon-Nucleon Collisions and the ABC
  Effect -\\ Approaching a Puzzle by
  Exclusive and Kinematically  Complete Measurements }
  
\author{ \Large T.~Skorodko$^1$, M.~Bashkanov$^1$, C.~Bargholtz$^{12}$, 
\Large D.~Bogoslawsky$^2$,\\
\Large H.~Cal\'en$^3$, 
\Large F.~Cappellaro$^4$,
 H.~Clement$^1$, L.~Demiroers$^5$, \\
\Large E.~Doroshkevich$^1$, D.~Duniec$^4$,
\Large C.~Ekstr\"om$^3$,
K.~Fransson$^3$, \\
\Large L.~Geren$^{12}$
\Large L.~Gustafsson$^4$,
B.~H\"oistad$^4$,
\Large G.~Ivanov$^2$,
M.~Jacewicz$^4$,\\
\Large E.~Jiganov$^2$,
T.~Johansson$^4$, 
\Large M.~Kaskulov$^1$,
O.~Khakimova$^1$,\\
\Large S.~Keleta$^4$,
\Large I.~Koch$^4$,
F.~Kren$^1$, 
\Large S.~Kullander$^4$,
A.~Kup\'s\'c$^3$,\\
\Large A.~Kuznetsov$^2$,
\Large K. Lindberg$^{12}$, \Large P.~Marciniewski$^3$,
B.~Martemyanov$^{11}$,\\ 
\Large R.~Meier$^1$,
\Large B.~Morosov$^2$,
\Large W.~Oelert$^8$,
\Large C.~Pauly$^5$, H.~Pettersson$^4$,\\
\Large Y.~Petukhov$^2$,
\Large A.~Povtorejko$^2$, A.~Pricking$^1$,
\Large R.J.M.Y.~Ruber$^3$, \\
\Large K.~Sch\"onning$^4$,
\Large W.~Scobel$^5$,
\Large B.~Shwartz$^{9}$,
\Large V.~Sopov$^{11}$,\\
\Large J.~Stepaniak$^7$,
\Large P.-E.~Tegner$^{12}$,
P.~Th\"orngren-Engblom$^4$,\\
\Large V.~Tikhomirov$^2$,
\Large A.~Turowiecki$^{10}$,
\Large G.J.~Wagner$^1$, 
\Large M.~Wolke$^4$,\\
\Large A.~Yamamoto$^6$,
\Large J.~Zabierowski$^{7}$,
\Large I.~Zartova$^{12}$,
J.~Z{\l}omanczuk$^4$
 \bigskip \\
{\it $^1$~Physikalisches Institut der Universit\"at T\"ubingen, D-72076
  T\"ubingen, Germany} \\
{\it $^2$~Joint Institute for Nuclear Research, Dubna, Russia} \\
{\it $^3$~The Svedberg Laboratory, Uppsala, Sweden} \\
{\it $^4$~Uppsala University, Uppsala,Sweden} \\
{\it $^5$~Hamburg University, Hamburg, Germany} \\
{\it $^6$~High Energy Accelerator Research Organization, Tsukuba, Japan} \\
{\it $^7$~Soltan Institute of Nuclear Studies, Warsaw and Lodz, Poland} \\
{\it $^8$~Forschungszentrum J\"ulich, Germany} \\
{\it $^{9}~$Budker Institute of Nuclear Physics, Novosibirsk, Russia} \\
{\it $^{10}$~Institute of Experimental Physics, Warsaw, Poland} \\
{\it $^{11}$~Institute of Theoretical and Experimental Physics, Moscow,
  Russia} \\ 
{\it $^{12}$~Department of Physics, Stockholm University, stockholm, Sweden} \\
~~~~~~~~ \\
{\it (CELSIUS-WASA Collaboration)}}

\maketitle

{\large

\begin{center}
{\bf Abstract}\\
\medskip

The ABC effect - 
a puzzling low-mass enhancement in the $\pi\pi$ invariant mass spectrum - is
known from inclusive measurements of two-pion production in nuclear
collisions, where it always showed up, if the participating nucleons fused to
a bound nuclear system in the final state. 
The first exclusive measurements on the ABC effect have been 
carried out very recently at CELSIUS-WASA for the fusion reactions leading to
d, $^3$He and $^4$He nuclei in the final state. The data analyzed so far for
the fusion processes to d and $^3$He reveal this effect to be a
$\sigma$ channel phenomenon associated with the formation of a strongly
attractive $\Delta\Delta$ system. The data for the strictly isospin-selective
double-pionic fusion to $^4$He, where we expect the 
largest effect, are currently still analyzed. All inclusive data on this system
are well described by our model, too. This case also constitutes the
heaviest nuclear system, where 
exclusive measurements of double-pionic fusion can be carried out with
present-day instruments. Surprisingly, the $pp \to pp\pi^0\pi^0$ reaction in
the $\Delta\Delta$ region is observed to also show a ABC-like low-$\pi\pi$
mass enhancement, a phenomenon, which deserves special attention.

\end{center}

\section{Introduction} \label{s1}
The ABC effect - first observed by Abashian, Booth
and Crowe \cite{abc} -  in
the double pionic fusion of deuterons and protons to $^3$He, stands for an
unexpected enhancement at low masses in the
$M_{\pi\pi}$ spectrum. Follow-up experiments \cite{ban,wur,col} revealed this
effect to be of isoscalar nature and to show up in cases, when the two-pion
production process leads to a bound nuclear system. With
the exception of low-statistics bubble-chamber measurements all
experiments conducted on this issue have been inclusive measurements carried
out preferentially with single-arm magnetic spectrographs for the detection
of the fused nuclei.

Initially the low-mass enhancement had been interpreted as an unusually large 
$\pi\pi$ scattering length and evidence for the $\sigma$ meson,
respectively. Lateron the ABC effect has been interpreted by $\Delta\Delta$
excitation
in the course of the reaction process leading to both a low-mass and a
high-mass enhancement in isoscalar $M_{\pi\pi}$ spectra. In fact, the
missing momentum spectra from inclusive measurements have been in support of
such predictions. It has been shown \cite{alv} that these structures can be
enhanced considerably in theoretical calculations including $\rho$ exchange.
In this case  even
the basic reaction $pp \rightarrow pp\pi\pi$ with no bound state in the exit
channel should exhibit a double-hump structure in the $M_{\pi\pi}$  spectrum
for incident energies in the $\Delta\Delta$ region.

\section{Exclusive Measurements at CELSIUS-WASA}

In order to shed more light on this issue exclusive
measurements of the reactions $pp \rightarrow NN\pi\pi$ (at several energies),
$pn \rightarrow d\pi^0\pi^0$ ($T_p$ = 1.1 GeV), $pd \rightarrow ^3$He$\pi\pi$
($T_p$ = 0.893 GeV) and $dd \rightarrow ^4$He$\pi\pi$ ($T_p$ = 1 GeV)
have been carried out in 
the energy region of the ABC effect at CELSIUS using the
4$\pi$ WASA detector setup including the pellet target system.

The $dd \rightarrow ^4$He$\pi\pi$ reaction could be measured only with modest
 statistics in the very last experimental run before the final shutdown of
the CELSIUS ring. The latter data are currently analyzed by the Uppsala group. 

For the other reactions quoted above there are
already first results. Those on the $pd \rightarrow
^3$He$\pi^0\pi^0$ and $pd \rightarrow ^3$He$\pi^+\pi^-$ reactions have been
published already \cite{bash}. As an example of the experimental
performance and the ability of particle identification we show in Fig.1 a
$\Delta$E-E scatterplot for particles registered in the forward detector of the
WASA-setup \cite{zab} in a run with proton beam and deuteron pellet target.

$^3$He particles, deuterons and protons have been detected in the forward
detector and identified by the $\Delta$E-E technique using corresponding
informations from quirl and range hodoscope, respectively. In order to
suppress the vast background of fast protons and other minimum ionizing
particles already on the trigger level, appropriate $\Delta$E thresholds have
been set on the window hodoscope acting as a first level trigger in case of
the $^3$He run. 

Charged pions and gammas (from $\pi^0$ decay) have been detected in the
central detector. This way the full four-momenta have been measured for all
particles of an event allowing thus kinematic fits with 4 overconstraints in
case of $\pi^+\pi^-$ production and 6 overconstraints in case of $\pi^0\pi^0$
production \footnote{In case of $pn \rightarrow d\pi^0\pi^0$, which 
has been measured as a quasifree reaction in $pd$ collisions, a kinematic fit
with only 3 overconstraints was applicable, since the
spectator proton had not been measured}.

\begin{figure}
\begin{center}
\includegraphics[width=0.70\textwidth]{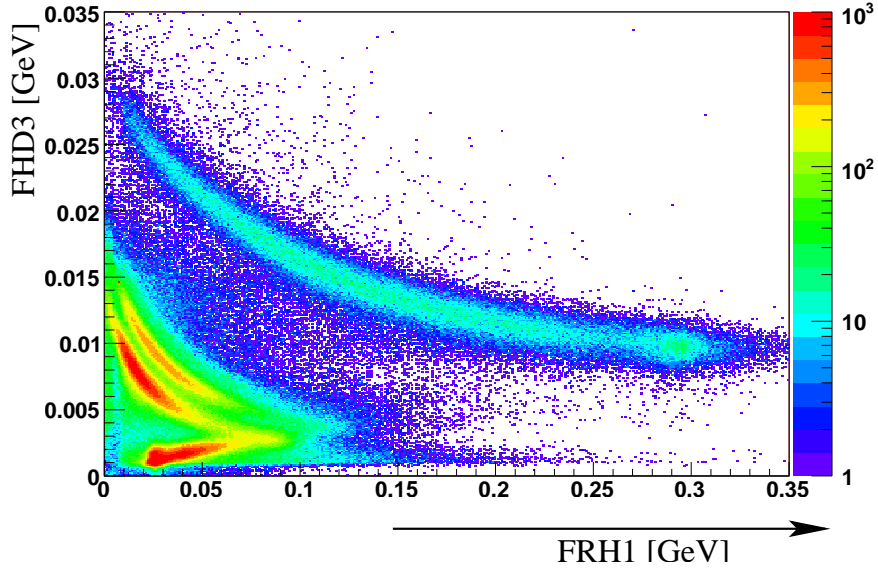} 
%\vspace*{8pt}
\caption{$\Delta$E-E scatterplot for particles recorded in the WASA Forward
    Detector during a run with proton beam and deuterium
    pellet target at $T_p$ = 0.895 GeV. The hyperbolic bands for $p$, $d$ and
    $^3$He particles (bottom to top) are clearly separated
    \protect\cite{MB}. Note the log scale in the scatterplot.}
\label{fig1}
\end{center}
\end{figure}

\begin{figure}
\begin{center}
%\vspace{3pt}
\includegraphics[width=18pc]{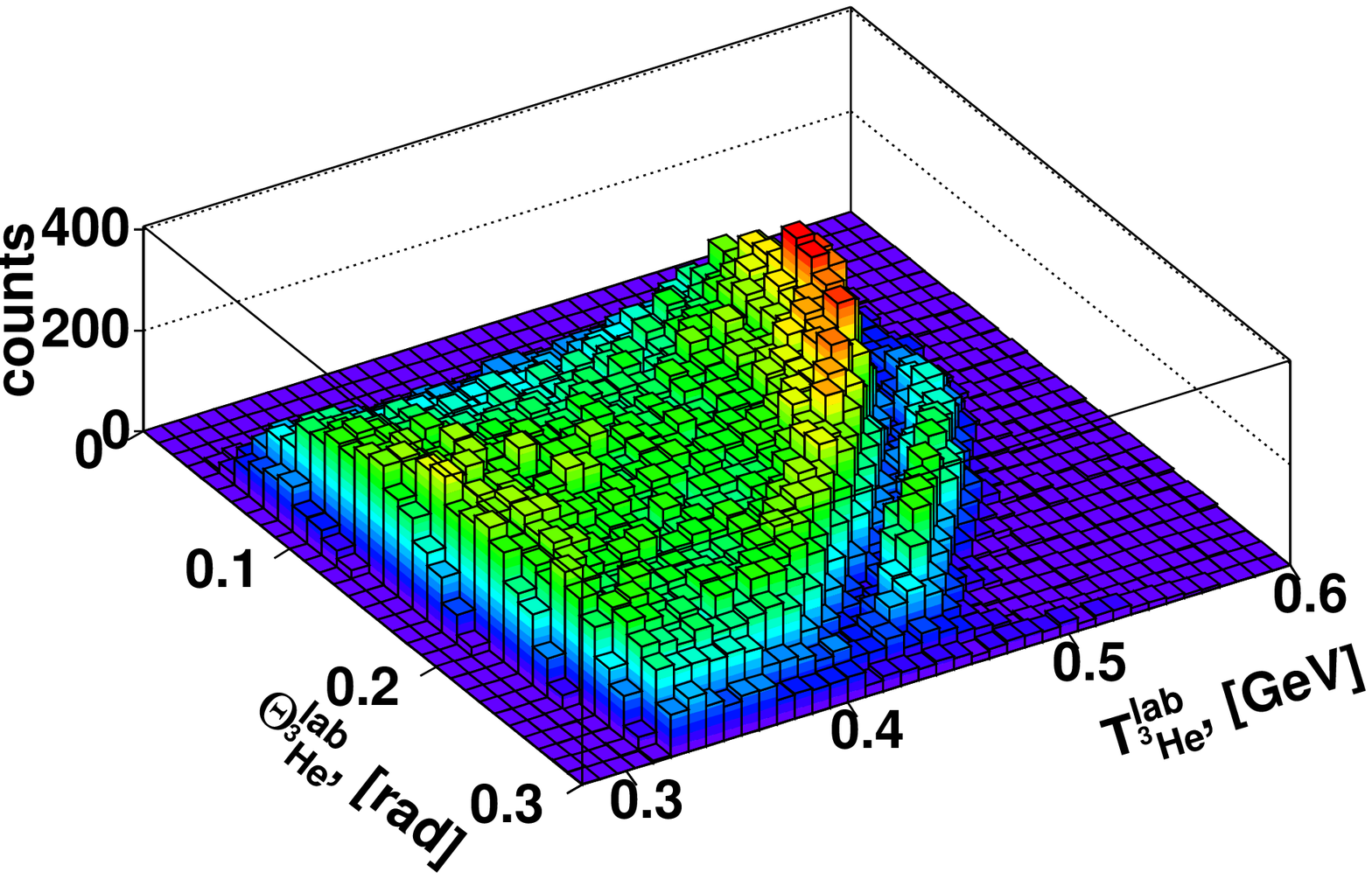}
\includegraphics[width=18pc]{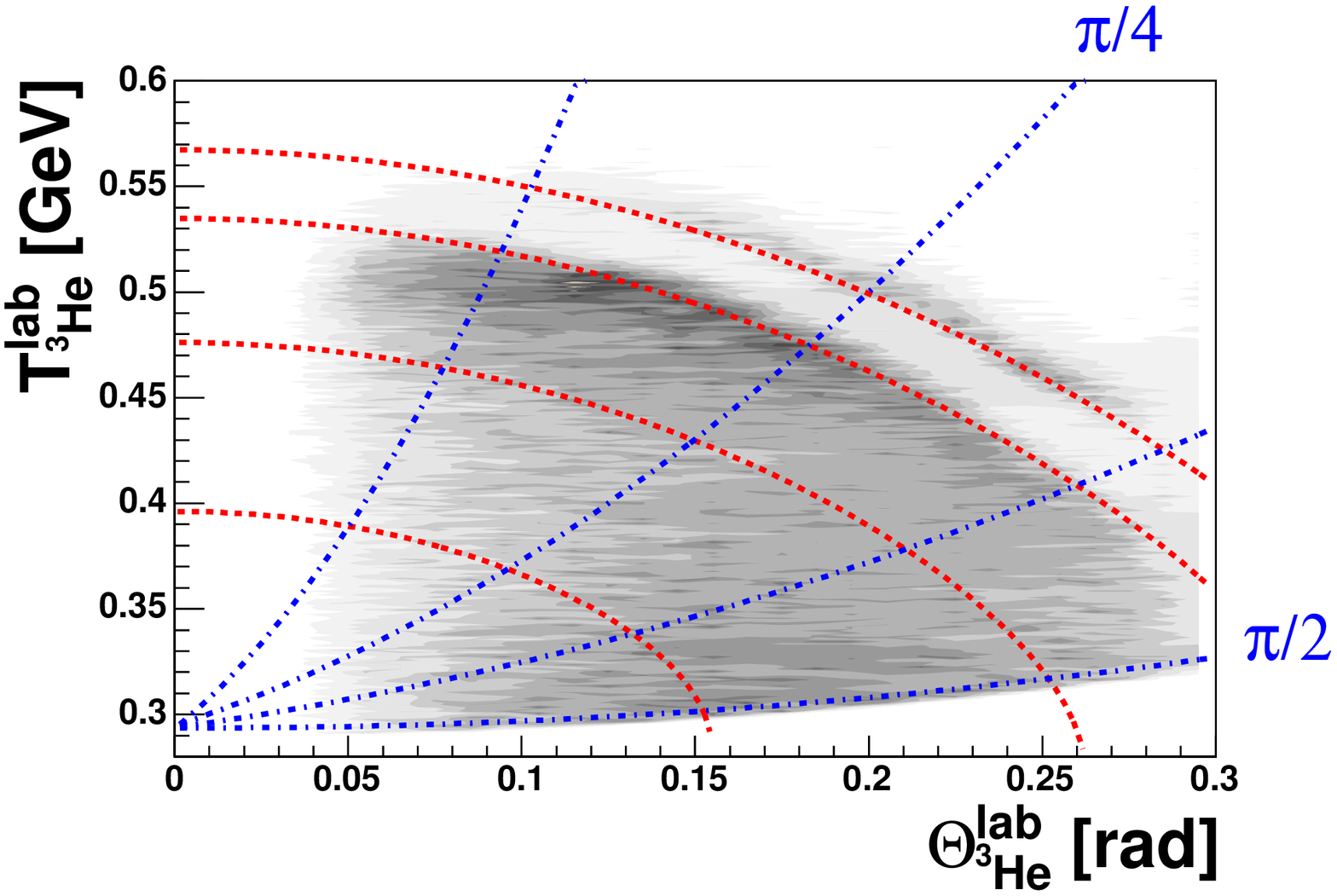}
\end{center}
\caption{3D and contourplots of lab angle $\Theta^{lab}_{^3He}$ versus lab
  energy $T^{lab}_{^3He}$for $^3He$ particles measured in the forward
  detector. The dash-dotted lines give $\Theta^{cm}_{^3He}=22.5^\circ,
  45^\circ,  67.5^\circ, 90^\circ$.  The dashed lines indicate the contours of
  missing masses $MM_{^3He} =$ 0.135, 0.27, 0.4, 0.5 GeV. From Ref. \cite{bash}.  }    
%\label{fig:largenenough}
%\label{fig:small}
\end{figure}

%\begin{figure}
%\begin{center}
%\includegraphics[width=24pc]{/home/pion/clement/paper/He3_notSelected_Data_models.eps}
%\end{center}
%\caption{$^3$He momentum spectrum for the angular bin  $7^\circ \leq
%\Theta^{lab}_{^3He} \leq 8^\circ$ (not corrected for detector efficiency). The
%data points represent the
%inclusively measured spectrum for comparison with the Saclay data
%\cite{ban}. The shaded histogram displays the phase space for $\pi\pi$
%production normalized as to touch the data as in Ref.\cite{ban}. The
%dash-dotted histogram shows the phase space calculation for $\pi\pi\pi$
%production. The dashed histogram represents conventional $\Delta\Delta$
%calculations normalized to the data, whereas the solid histogram shows the same
%calculations with a boundstate condition for the $\Delta\Delta$ system
%included. From Ref. \cite{bash}.
%} 
%\end{figure}

In order to facilitate comparison with the previous inclusive measurements
\cite{ban} we
display in Fig. 2 lego and contour plots of lab angle versus lab energy of
the $^3$He particles detected in the forward detector - before kinematic fit
and any demand on other particles in the event. Whereas for single $\pi^0$
production $^3$He particles have been registered only in a very limited angle
and energy range of phase space, the $^3$He particles stemming from $\pi\pi$
production have been detected over the full kinematical range
up to $^3$He angles $\Theta^{cm}_{^3He} \leq 90^\circ$. Since we demand the
$^3$He particles to reach the range hodoscope they need to have kinetic
energies of more than 200 MeV in order to be registered and safely
identified. Hence for $^3$He cms angles larger than $90^\circ$ the phase
space is no longer fully covered in this measurement.

In Fig. 2 the band for
single $\pi^0$ production is seen to be well separated from the continuum for
$\pi\pi$ production. Also immediately evident is a large accumulation of
events near the kinematic limit for $\pi\pi$ production, i.e. in the region
corresponding to small invariant $\pi\pi$ masses. Since the detector efficiency
is approximately constant over the corresponding phasespace region in Fig.2,
this feature obviously is in accord with a strong ABC enhancement present in
these data.

\section{Results from Exclusive Measurements at CELSIUS-WASA}

The $\pi^0\pi^0$ channel, which is free of any isospin I=1
contributions, exhibits in all cases a low-mass enhancement in the
$M_{\pi\pi}$ spectra. In Fig. 3 the situation is shown for the
reactions  $pd \rightarrow ^3$He$\pi^0\pi^0$ and $pd \rightarrow
^3$He$\pi^+\pi^-$ at $T_p$ = 0.895 GeV. The chosen energy is in the region,
where the ABC effect is expected to be at maximum as deduced from the inclusive
measurements on this reaction system\cite{ban}. For the  $\pi^0\pi^0$
channels associated with a bound deuteron \cite{OK} or an unbound $pp$ system
\cite{TS} in the 
final nuclear system the situation is similar, see Figs. 4 and 7.

\begin{figure}[t]
\begin{center}
\includegraphics[width=22pc]{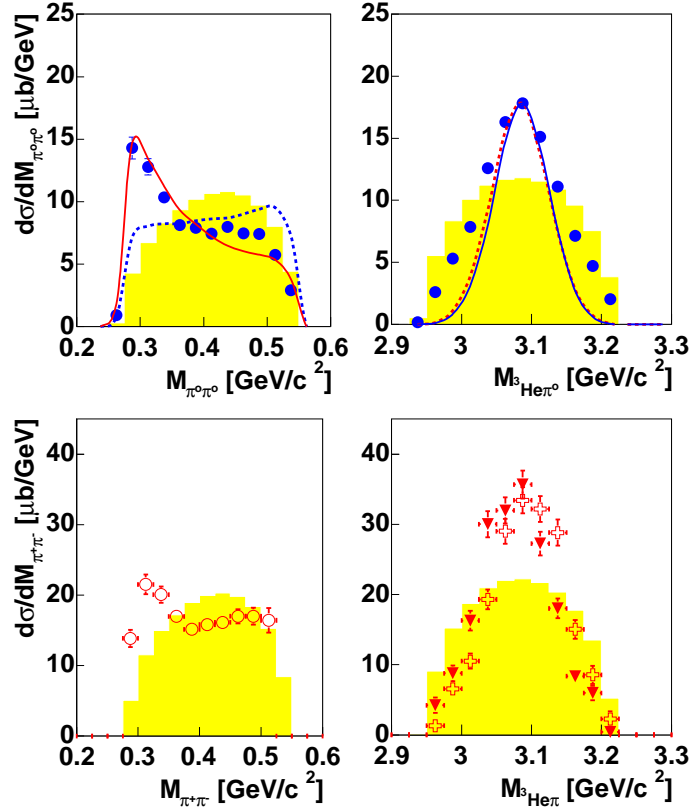}

%\vspace*{8pt}
\caption{Results from the exclusive measurements of the reactions $pd
  \rightarrow ^3$He$\pi^0\pi^0$  ({\bf top}) and $pd \rightarrow
  ^3$He$\pi^+\pi^-$  ({\bf bottom}) at $T_p$ = 0.895 GeV. 
  {\bf Left}: Spectra of the $\pi\pi$ invariant mass $M_{\pi\pi}$.
  {\bf Right}: Spectra of the $^3$He$\pi$ invariant masses $M_{^3He\pi^0}$
  (solid dots),
  $M_{^3He\pi^+}$ (open crosses) and  $M_{^3He\pi^-}$ (filled triangles). The
  shaded areas show the pure phase
  space distributions ( left: normalized to touch the data; right: normalized
  to the total cross section of the data. Solid and dashed curves give
  $\Delta\Delta$ calculations with and without $\Delta\Delta$ interaction,
  respectively \protect\cite{bash,MB}. Note that in these calculations no
  collison damping is taken into account, which may the reason, why the
  observed structure in the $^3$He$\pi$ invariant masses is larger than the
  calculated $\Delta$ width.}
\label{fig3}
\end{center}
\end{figure}

\begin{figure} 
\begin{center}
\includegraphics[width=14pc]{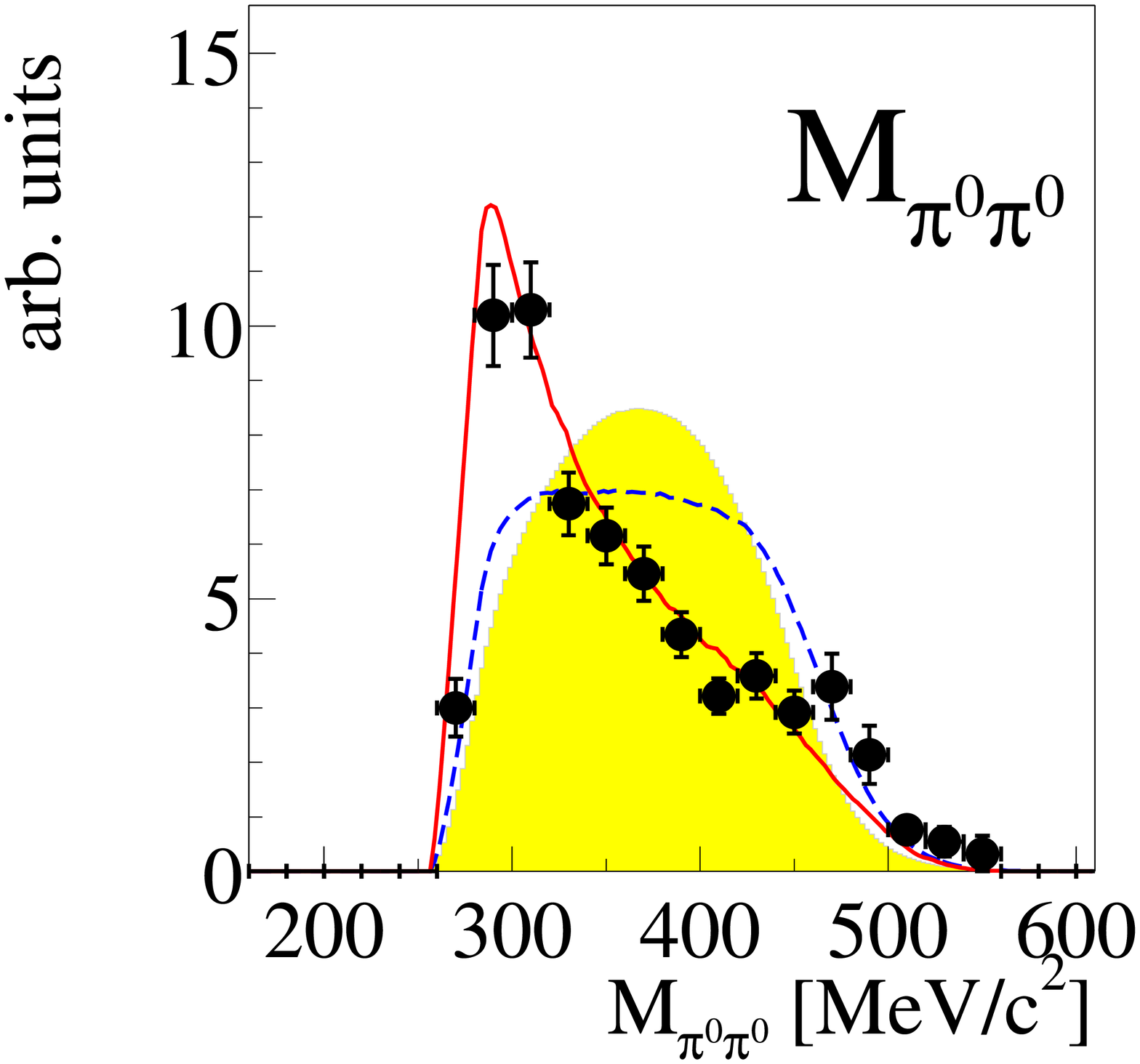}  
\includegraphics[width=14pc]{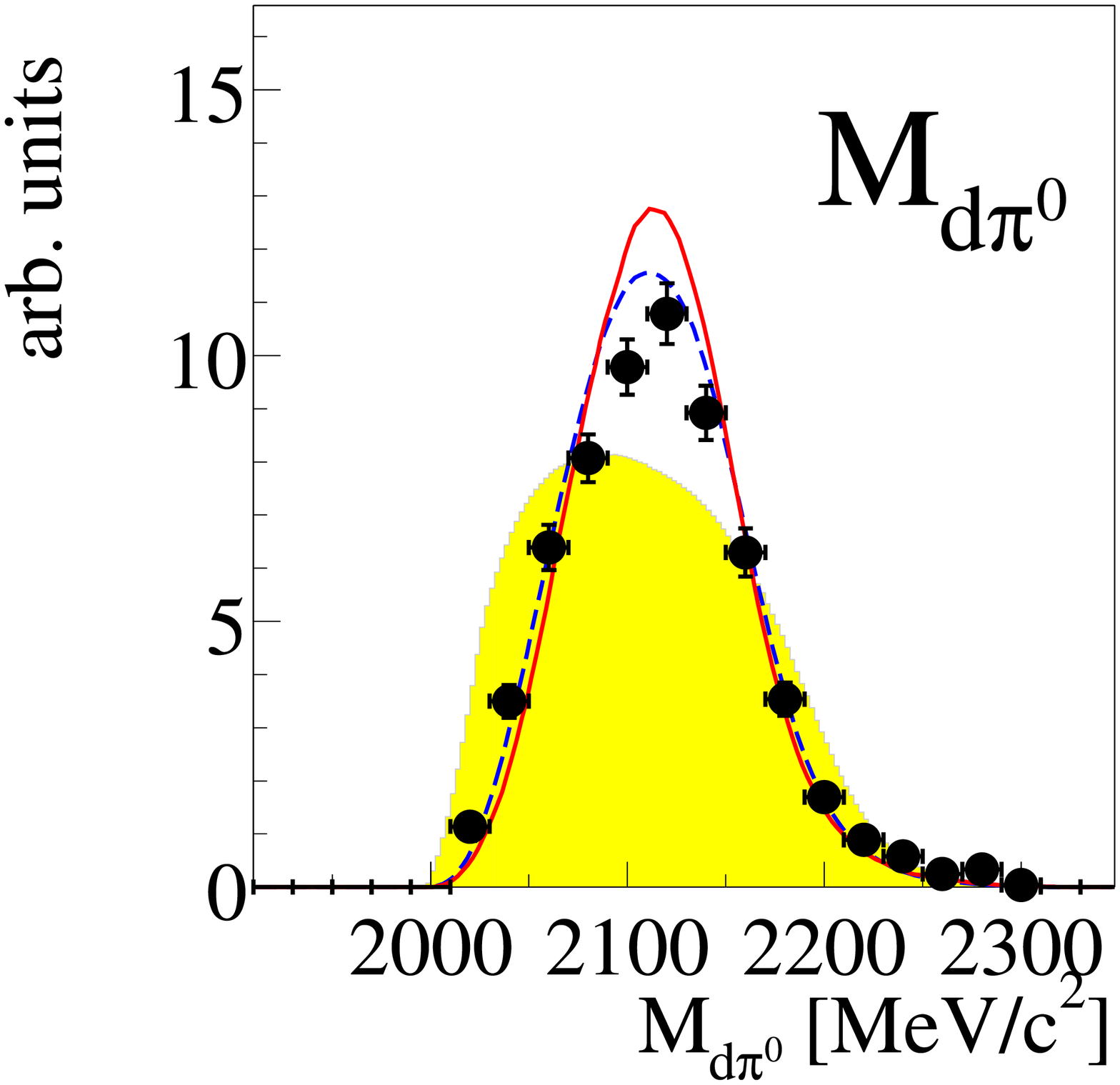} 
\caption{Preliminary results\cite{OK} from exclusive measurements of the 
  reaction $pn\rightarrow d\pi^0\pi^0$  at $T_p$ = 1.05 GeV. 
  {\bf Left}: Spectrum of the $\pi\pi$ invariant mass $M_{\pi^0\pi^0}$.
  {\bf Right}: Spectrum of the $d\pi$ invariant mass $M_{d\pi^0}$. The
  shaded areas show the pure phase space distributions, solid and dashed
  curves give $\Delta\Delta$ calculations with and without $\Delta\Delta$
  interaction, respectively\protect\cite{MB}. Note that in this case no
  collision damping is expected to occur. Hence the calculated width in the
  $M_{d\pi^0}$ spectrum should fit to the observed one - as indeed is the case.}
\label{fig4}
\end{center}
\end{figure}

For the $\pi^+\pi^-$ channel, which may
contain isovector contributions, the observed threshold enhancements show up
much less pronounced in case of $^3$He and even are absent in case of the
unbound $pp$ system in the final state.

From the angular distribution in the $\pi\pi$ subsystem we see that the 
threshold enhancement is of scalar nature\cite{bash,MB}. 
%As an example angular
%distributions are shown in Fig. 5 for the reactions  $pd \rightarrow
%^3$He$\pi^0\pi^0$ and $pd \rightarrow 
%^3$He$\pi^+\pi^-$ at $T_p$ = 0.895 GeV. 

The $\pi\pi$ low-mass enhancements observed now in the exclusive data for the
$\pi^0\pi^0$ channels turn out to be much
larger than predicted in previous $\Delta\Delta$
calculations\cite{ris,anj,gar}. At the same time the data do not exhibit any
high-mass enhancement as predicted by the same calculations and as suggested by
the inclusive measurements. As anticipated already in Ref.\cite{col} the
high-mass bump observed in inclusive spectra appears to be associated with
$\pi\pi\pi$ production and I=1 contributions rather than with the isoscalar
$\pi\pi$ production. 

%\begin{figure} 
%\begin{center}
%%\vspace{3pt}
%\includegraphics[width=30pc]{/home/pion/clement/paper/new_fig_ang1.eps}
%\end{center}
%\caption{Angular distributions of  
% the opening angle $\delta_{\pi\pi}$ between the two pions, the angle of the
% total momentum of the 
% $\pi\pi$ system $\Theta^{cm}_{\pi\pi}=-\Theta^{cm}_{^3He}$ - all in
% the overall cms - as well as the pion angle $\Theta^{\pi\pi}_{\pi}$ in the
% $\pi\pi$ subsystem  (Jackson frame). For the latter the data 
% are plotted also 
% with the  constraint $M_{\pi\pi} < 0.34$ GeV (open squares). {\bf Top}:
%  $pd\rightarrow$ $^3$He 
%  $\pi^0\pi^0$, {\bf bottom}: $pd\rightarrow$ $^3$He $\pi^+\pi^-$. For the
% meaning of 
% symbols and curves see caption of Fig. 3. From Ref. \cite{bash}}
%%\label{fig:largenenough}
%\label{fig:toosmall}
%\end{figure}

Since on the one hand the available $\Delta\Delta$ calculations
obviously fail, but on the other hand the data clearly show the $\Delta\Delta$
excitation in their $M_{N\pi}$ spectra, a profound physics piece appears to be
missing. Such a missing piece is provided by a strong $\Delta\Delta$
attraction, which is able to describe the exclusive data for d and $^3$He
fusion as well as the inclusive spectra for double-pionic fusion to $^4$He
amazingly well (see Figs. 5 and 6) without any change of the
$\Delta\Delta$ interaction parameters \cite{MB}.

\begin{figure}[t]
\begin{center}
\includegraphics[width=20pc]{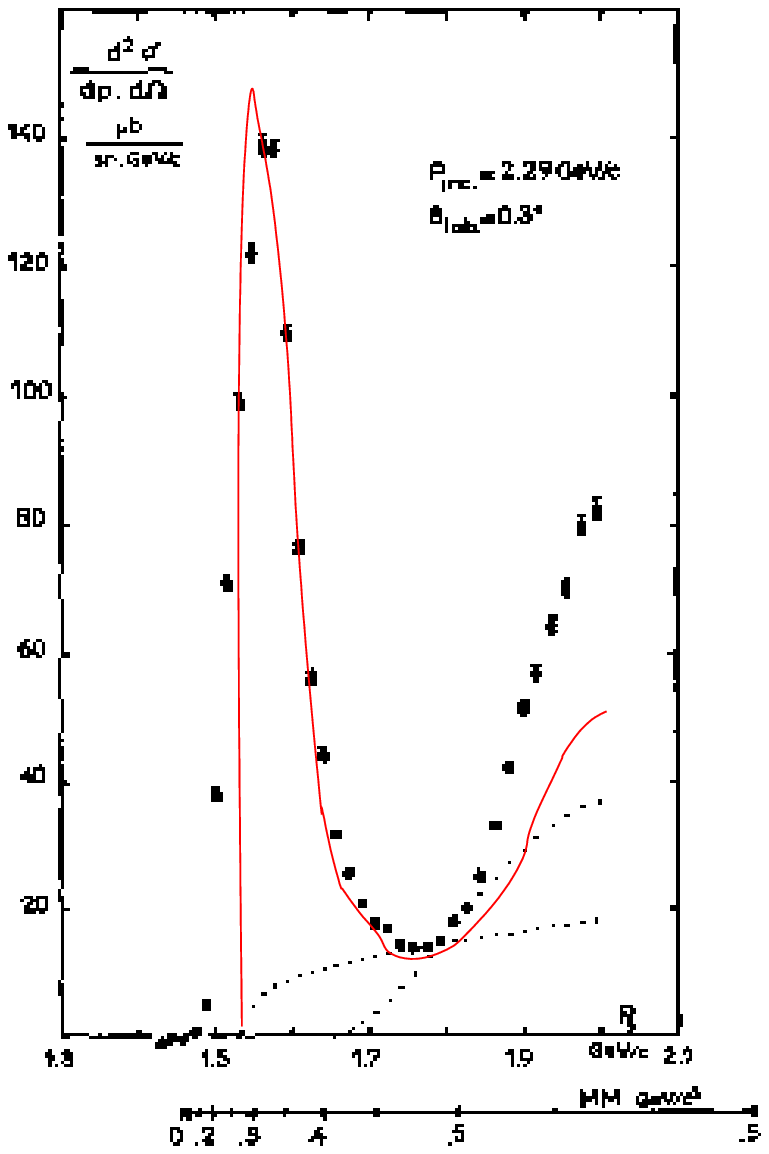}
\end{center}
\caption{ Experimental $^4$He momentum spectrum for the reaction $dd \to ^4$He
  X at $T_d$ = 1.09 GeV  and $\Theta_{lab}$ = 0$^\circ$ as obtained at Saclay
\cite{ban1}. The solid (red) line displays our prediction with
$\Delta\Delta$ interaction (adjusted in height to the data), whereas the
broken lines are from the original paper and denote phase space distributions
from for two- and three-pion production - each adjusted to touch the data. From
Ref. \cite{MB}}
\end{figure}

\begin{figure}
\begin{center}
\includegraphics[width=26pc]{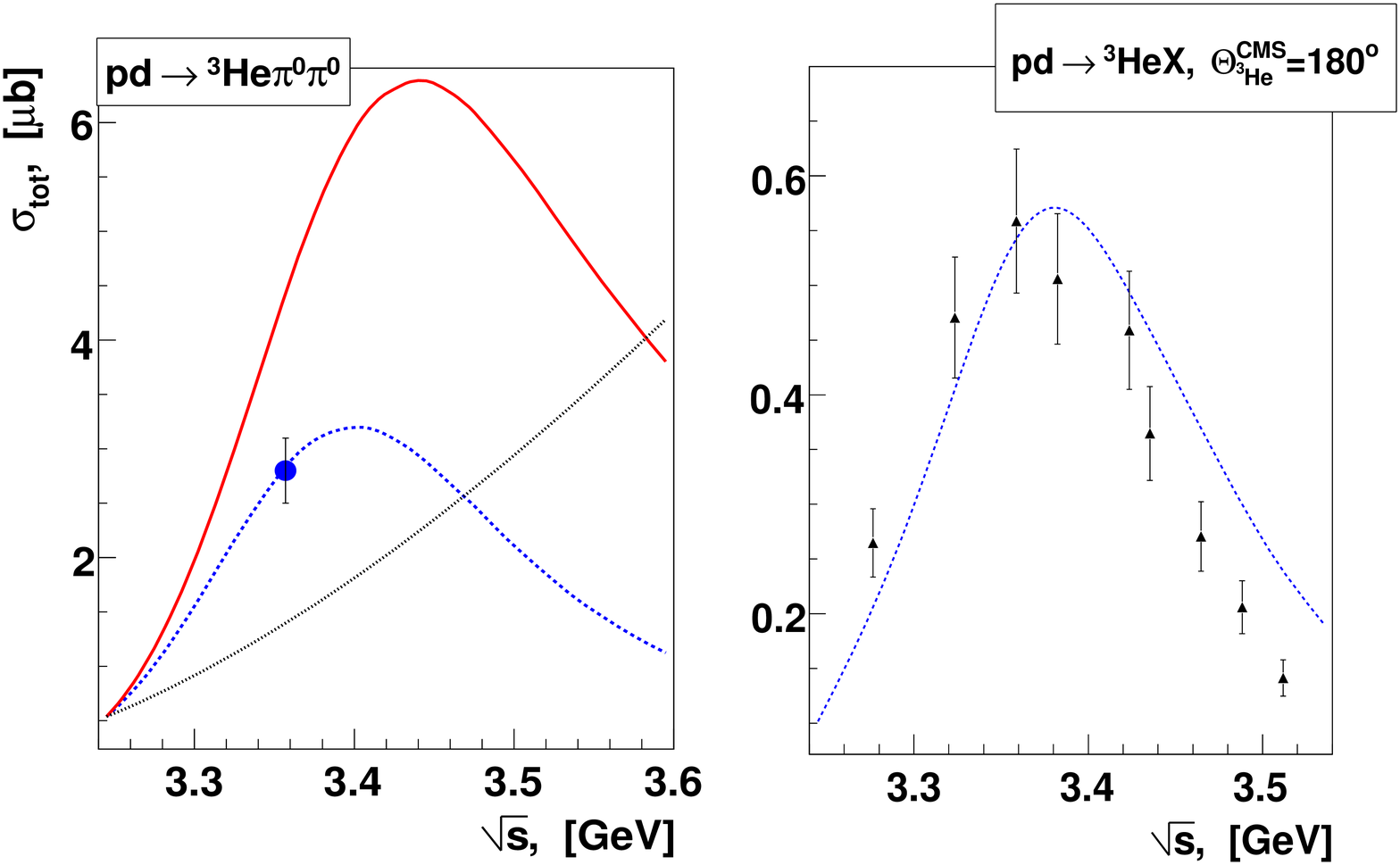}
\end{center}
\caption{Energy dependence of double pionic fusion, shown here for the
  reaction $pd \to ^3$He $\pi^0\pi^0$ as an example. The situation for the
  $^4$He case is very similar. The dotted (black) curve
  denotes phase 
  space, dashed (blue) and solid (red) curves are predictions of
  $\Delta\Delta$ excitation with
  and without $\Delta\Delta$ interaction included. {\bf Left}: total cross
  scetion with the solid dot denoting the
  CELSIUS-WASA result \cite{bash} and the blue curve being arbitrarily
  adjusted to it. {\bf Right}: ABC peak, i.e. $\pi\pi$ low-mass enhancement
  above 
  phase space (where the latter is adjusted to touch the data, see e.g.
  Figs. 5) at $\Theta_{^3He}^{cm}$ = 180 $^\circ$. Comparison of data
  taken from 
  \cite{ban1} with predictions of the $\Delta\Delta$ ansatz with interaction
  included. From Ref. \cite{MB}} 
\end{figure}

The double-pionic fusion of deuterium to $^4$He constitutes the fusion process
to the heaviest nucleus, which can be measured exclusively with
present-day experimental setups. For exclusive measurements on heavier nuclei
high-resolution zero-degree spectrometers directly connected to the vacuum and
supplemented by a WASA-like central detector would be required - an
instrument, which is not available at present.

The fusion process to $^4$He exhibits the biggest ABC effect ever
measured in previous inclusive magnetic spectrometer measurements \cite{wur}.
The $^4$He momentum spectrum, which represents the $\pi\pi$ invariant mass
spectrum (though in a nonlinear way), decomposes in this case in nearly two
narrow bumps corresponding to low- and high-mass $\pi\pi$-enhancements - see
Fig. 5, where the experimental momentum spectrum as obtained at Saclay
\cite{ban} is shown. The solid line displays our prediction with
$\Delta\Delta$ interaction (adjusted in height to the data), whereas the
broken lines are from the original paper \cite{ban1} and denote phase space
distributions for two- and three-pion production each adjusted to touch the
data. The prediction compares very well with the data with the exception of
the high-mass region, where we suspect strong contributions from three-pion
production in the inclusive spectra.

Yet another prediction is the energy dependence of the reaction cross section,
which should be resonance-like. This is demonstrated in Fig. 6, which on the 
left shows the
situation for the total cross section and on the right the situation of the
ABC effect itself. As in the original work on $^4$He \cite{ban1} the latter is
taken to be the $\pi\pi$ low-mass enhancement above phase space, where the
latter is adjusted as to touch the data (as displayed in Fig. 5). Again our
prediction does remarkably well.

The lecture we are learning here for the $\pi\pi$ production in
few-nucleon systems may possibly serve as a guideline for the microscopic
understanding 
of medium effects in the $\sigma$ channel and its possible association to
chiral restoration.

%\begin{figure}
%\begin{center}
%\includegraphics[width=22pc,angle=-90]{/home/pion/clement/proposal/Frh1vFrh2.eps}
%\end{center}
%\caption{$\Delta$E-E plot from the very last CELSIUS-WASA run June 2005 with
%  deuteron beam on deuteron pellet target at $T_d$ = 1.012 GeV. The $\Delta$E
%  Signal is taken from FRH1 ( Forward Range Hodoscope layer 1), whereas the E
%  signal is from FRH2. From Ref. \cite{SK}} 
%\end{figure}

\section{The Basic Reaction Leading to Unbound Final States}

For the basic reaction $pp \to NN\pi\pi$  most of the data have been 
collected for the $\pi^+\pi^-$  and $\pi^0\pi^0$ channels, which are most
easily accessible experimentally. The $\pi^+\pi^+$  and $\pi^+\pi^0$ channels
are much harder to access, since they are associated
with the production of neutrons.

\begin{figure}[t] 
\begin{center}
\includegraphics[width=22pc]{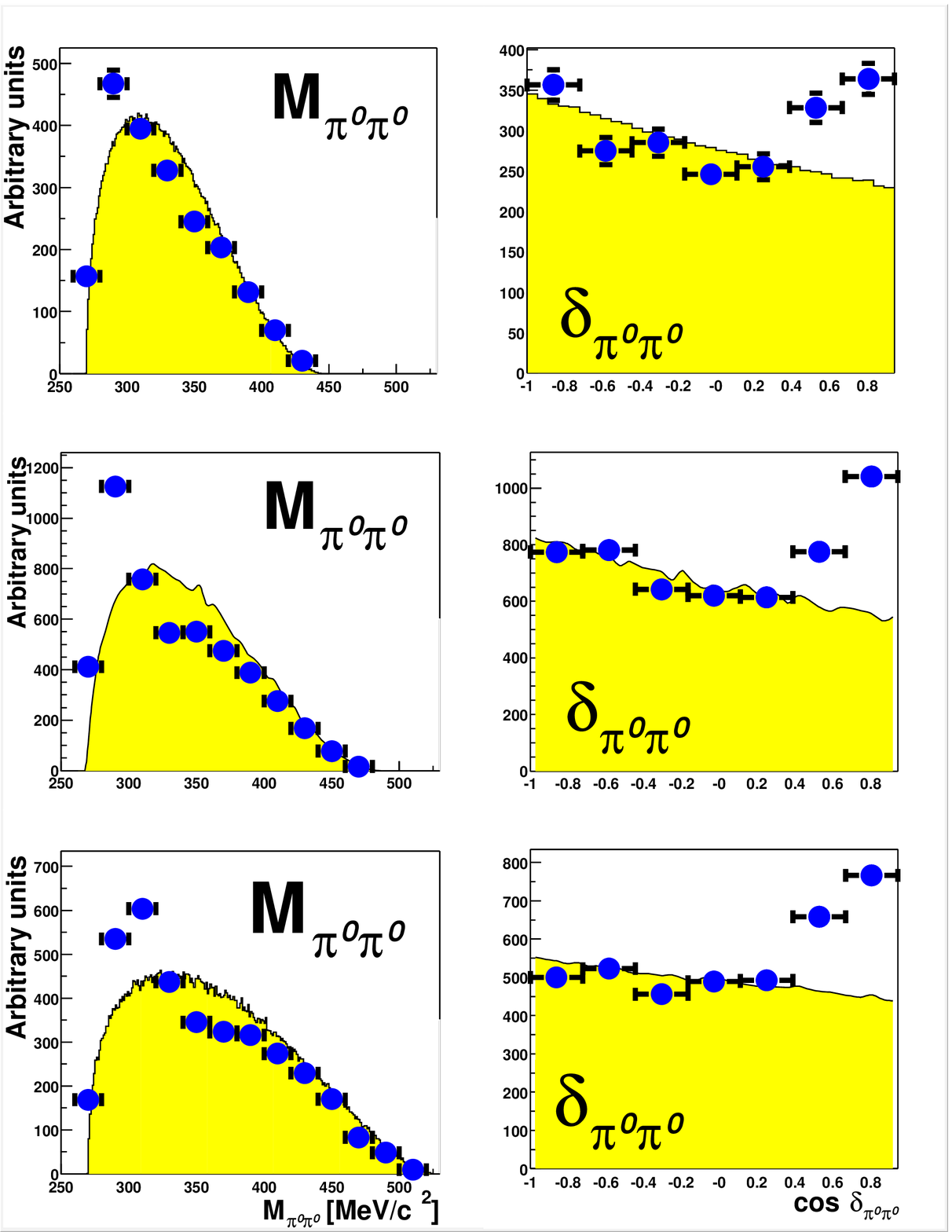}  
%\vspace*{8pt}
\caption{Preliminary results\cite{TS} from the exclusive measurements of the 
  reaction $pp\rightarrow pp\pi^0\pi^0$  for 
  $\pi^0\pi^0$ invariant masses $M_{\pi^0\pi^0}$ ({\bf left}) and
  opening angles $\delta_{\pi^0\pi^0}$ between the two pions in
  the overall center-of-mass system ({\bf right})
 at $T_p$ = 1.0 GeV ({\bf top}), 1.1 GeV  ({\bf middle}) and 1.2 GeV {\bf
 bottom}). The shaded areas show the corresonding distributions for pure phase
  space.
  }
\label{fig8}
\end{center}
\end{figure}

At low incident energies, i.e., in the threshold region, the data on the
$\pi^+\pi^-$  and $\pi^0\pi^0$ channels have been successfully explained by
excitation and decay of the Roper resonance \cite{alv,bro,pae} - see also 
the corresponding article \cite{hcl} in this proceedings - and
alternatively  also within the concept of chiral dynamics \cite{kas}.

At incident energies above 1 GeV, where the $\Delta\Delta$ mechanism should
take over, the data for $\pi^+\pi^-$  and $\pi^0\pi^0$ channels change
drastically. Indeed this mechanism is identified by observing the
simultaneous excitation of $\Delta^{++}$ and $\Delta^{0}$ in the
appropriate $M_{p\pi^+}$ and  $M_{p\pi^-}$ spectra. However, we
observe\cite{slov} a phase-space like behavior in the measured
$M_{\pi^+\pi^-}$ spectra rather than the predicted\cite{alv} double-hump
structure. This observation together with the observation that the measured
pion angular 
distributions are flat has led us to the conclusion that obviously the
$\Delta\Delta$ system is produced solely in an $I(J^P)$ = $1(0^+)$
configuration . This is
equivalent to the statement that the two-pion production
process is dominantly fed by the $^1S_0$ partial wave in the
entrance channel. This in turn is consistent with the observation at lower
energies, where the there dominating Roper process proceeds via
the $^1S_0$ incident partial wave, too. 
 
In the measured $M_{\pi^0\pi^0}$ spectrum on the other hand we find a
systematic low-mass enhancement, which is associated with the two pions
flying in parallel both in lab and overall center-of-mass systems - as
exhibited in the distributions of the $\pi^0\pi^0$ opening angle in Fig.7.    
This situation is in line with the situation observed
in the exclusive measurements of double-pionic fusion processes discussed
above for the ABC effect.

We note, however, that in the $pp$ collision case, 
where the nucleons in the final state stay unbound, the introduction of a
$\Delta\Delta$ interaction does not
produce any significant $\pi\pi$ low-mass enhancement. For a proper
description of the $\pi^0\pi^0$ low-mass enhancement there we possibly have to
go back to the original idea
\cite{abc} of an unusally strong $\pi\pi$ interaction as
the origin of the ABC effect. However, as we see only now from the
kinematically 
complete data samples, this unusual $\pi\pi$ interaction cannot be understood
to be the same one as studied in the vacuum but rather as a $\pi\pi$
interaction, which obviously is associated with the presence of nearby baryons
forming a $\Delta\Delta$  system in the intermediate state.

Yet another interpretation discussed previously in connection with $\pi\pi$
low-mass enhancement and ABC effect, respectively, has been the hypothesis of
Bose-Einstein 
correlations\cite{abc,slov}, which are expected to show up in case of sources
emitting identical bosons in a stochastically independent manner. However, in
such a scenario not only the  $\pi^0\pi^0$ channel should exhibit a low-mass
enhancement, but the  $\pi^+\pi^+$ channel, too. The latter channel is very
hard to access, since it is connected with the production of two neutrons, the
detection of which is very difficult experimentally. Using the cabability of
the WASA setup to detect neutrons with some finite efficiency, we have been
able to obtain exclusively measured data for the first time for this
channel. As a result no special $\pi^+\pi^+$ low-mass enhancement is found
there, excluding thus the Bose-Einstein correlation scenario \cite{TS}. 

So whereas the
double-pionic fusion reactions appear to have found a reasonable explanation
by the strong attractive force in the $\Delta\Delta$ system, this explanation
is not sufficient for the low-mass enhancement in the basic reaction leading
to the unbound two-nucleon system, for which an explanation is still pending.
However, one can not exclude that such an explanation, if found, might also
pertain to fused systems and thus provide an explanation for those as well,
providing an alternative to the one presented here.

\section{Acknowledgements}

We acknowledge valuable discussions with E. Oset, C. Wilkin, Ch. Hanhart,
A. Sibirtsev, A. Kaidalov, V. Pugatch and L. Dakhno on this issue.
This work has been supported by BMBF
(06TU201, 06TU261), COSY-FFE, 
DFG (Europ. Graduiertenkolleg 683), Landesforschungsschwerpunkt
Baden-W\"urttemberg and the Swedish Research Council. We also acknowledge the
support from 
the European Community-Research Infrastructure Activity under FP6
"Structuring the European Research Area" programme (Hadron Physics, contract
number RII3-CT-2004-506078).

\end{document}
%%%%%%%%%%%%%%%%%%%%%%%%%%%%%%%%%%%%%%%%%%%%%%%%%%%%%%%%%%%%%%%%%%%%%%%%%%%%%%%%%%%%%%%%%%%%%%%%%%%%%%%%%%%%%%%%%%%%%%%%%%%%%%%%%%%%%%%%%%%%%%%%%%%%%%%%%%%%%%%%%%%%%%%%%%%%%%%%%%%%%%%%%%%%%%%%%%%%%%%%%%%%%%%%%%%%%%%%%%%%%%%%%%%%%%%%%%%%%%%%%%%%%%%%%%%%%%%%%%%%%%%%%%%%%%%%%%%%%%%%%%%%%%%%%%%%%%%%%%%%%%%%%%%%%%%%%%%%%%%%%%%%%%%%%%%%%%%%%%%%%%%%%%%%%%%%%%%%%%%%%%%%%%%%%%%%%%%%%%%%%%%%%%%%%%%%%%%%%%

}%end of \large

\end{document}